\documentclass[EPJ,twocolumn]{webofc}
\usepackage[varg]{txfonts}   
\newcommand{\avXemmax}{$\left< {X}^{e.m.}_{max} \right>$}

\newcommand{\RMSXemmax}{$RMS(X^{e.m.}_{max})$}

\newcommand{\Xmax}{$X_{max}$}
\newcommand{\Xemmax}{$X^{e.m.}_{max}$}
\newcommand{\Xmumax}{$X^{\mu}_{max}$}

\newcommand{\qII}{QGSJ{\textsc{et}}-II}
\newcommand{\Conex}{$C{\textsc{onex}}$}

\newcommand{\beq}{\begin{equation}}
\newcommand{\eeq}{\end{equation}}
\wocname{ISVHECRI 2012}
\woctitle{ISVHECRI 2012}
\begin{document}
\title{The interplay between the electromagnetic and the muonic longitudinal profile at production}
%
%

\author{Ruben Concei\c{c}\~ao\inst{1}\fnsep\thanks{\email{ruben@lip.pt}} \and
       Sofia Andringa\inst{1}  \and
       Lorenzo Cazon\inst{1} \and
       M\'ario Pimenta\inst{1,2}
}

\institute{LIP, Av. Elias Garcia, 14-1, 1000-149 Lisboa, Portugal
\and
           Departamento de F\'{i}sica, IST, Av. Rovisco Pais, 1049-001 Lisboa, Portugal 
          }

\abstract{%
The electromagnetic and the muonic longitudinal profile at production enclosure important information about the primary particle and the hadronic interactions that rule the shower development. In fact, these two profiles provide two different insights of the shower: the electromagnetic component gives a measurement of the energy and the strength of the neutral pion channel; while the muonic profile, being intimately related with the charged mesons decays, can be used as a direct probe for the high energy hadronic interactions.

In this work we explore the interplay between the electromagnetic and muonic profiles, by analysing their phenomenologic behaviour for different primary masses and energies, zenith angles, and also different high energy hadronic interaction models. We have found that the muonic longitudinal profile at production displays universal features similar to what is known for the electromagnetic one. Moreover, we show that both profiles have new primary mass composition variables which are fairly independent of the high energy hadronic interaction model.

Finally we discuss how the information in the electromagnetic and the muonic longitudinal profile can be combined to break the degeneracy between the primary mass composition and the high energy hadronic physics.
}
\maketitle
\section{Introduction}
\label{sec:intro}
The Ultra High Energy Cosmic Rays (UHECRs) are the most energetic particles known. And yet its origin and composition remains a mystery although they were discovered half a century ago. The reason for this resides on the difficulty of their detection. Since their flux at the highest energies is very low the only way to detect them is by observing the huge cascades of secondary particles, known as  Extensive Air Showers (EAS), that are created by the interaction of these cosmic rays with the atmosphere molecules. 

The EAS can be detected by sampling the charged particles that arrive at the ground or, in moonless nights, the shower longitudinal profile can be sampled through the collection of the fluorescence light produced as the shower develops into the atmosphere. The latter measurement, although with a duty cycle of around 10\%, provides a direct measurement of the electromagnetic shower component development.

While the arrival direction of the UHECRs can be easily obtained by the experimental techniques aforementioned, the determination of the nature of the primary particle is much more difficult. In fact, the shower observables connected to the type of particle are also sensitive to the physical interactions that rules the shower development. Moreover, the hadronic interactions at high energies are described through phenomenological models that are fitted to the available accelerator data and extrapolated several orders of magnitude to the UHECRs energies. This implies that the determination of the UHECRs mass composition is linked to the understanding of the shower physical mechanisms, in particular to the hadronic interactions descriptions.

Muons are a fundamental tool to assess the high energy hadronic interaction models as they are produced essentially from charged mesons (secondary products from hadronic interactions). Moreover, a large fraction of the muons can reach the ground due to a low interaction cross-section and a relatively large decay time. Recent analyses, that use the muon arrival time, are able to reconstruct the muon longitudinal profile at production at ultra high energies~\cite{MPDAuger}, providing an additional insight of the shower. 

The phenomenological behaviour of the muonic longitudinal profile at production and its relation with the electromagnetic shower profile is the main focus of this work. Moreover, we show, at the end of this paper, how several shower observables may be combined to disentangle the mass composition from the hadronic interaction models.

The paper is organised as follow: in section~\ref{sec:long} the muonic longitudinal profile at production is presented and the main observables, namely \Xmax, are discussed. Whenever relevant the features of this profile will be compared to the ones found in the electromagnetic profile; In section~\ref{sec:usp} we show that the muonic profile shape exhibits an universal behaviour and its correlation with the shape of the electromagnetic profile is assessed; In section (sec.~\ref{sec:umbrella}) we show how hadronic interaction models can be distinguish independently of primary mass composition scenarios, through the combination of different shower observables. Finally, some conclusions are drawn in section~\ref{sec:conclusions}.

\section{Longitudinal shower profiles}
\label{sec:long}


To study the \emph{true}\footnote{the experiments measure the \emph{apparent}-MPD, which depends on the point of observation, as the true distribution has to be convoluted with the muons propagation. By its turn, the muon propagation depends of the distance travelled by the muons, their energy, among others. The transformation of the \emph{apparent}-MPD into the \emph{true}-MPD is described in~\cite{TransportModel}.} Muonic Production Depth profile (onwards referred as MPD profile) we have used \Conex~\cite{Conex2} simulations (for more details we refer the reader to~\cite{USPVmuons}).

\begin{figure}
\centering
\sidecaption
\includegraphics[width=0.45\textwidth,clip]{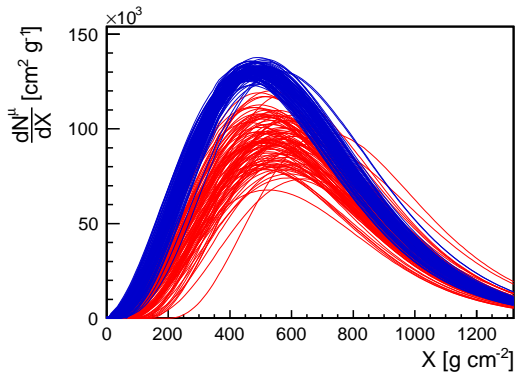}
\caption{Muon production shower profiles as a function of depth ($X$), for proton (red) and iron (blue) primaries at $E = 10^{19}$ eV. In this picture $100$ showers for each primary are shown, both generated with \qII.03 and with $\theta = 40^\circ$.}
\label{fig:prof}       
\end{figure}

The MPD\footnote{note that in \Conex\ there is an energy threshold for muons of $1$ GeV.} profile for proton induced showers (in red) and iron primaries (in blue) at $E=10^{19}$ eV is shown in Fig.~\ref{fig:prof}. Compared to proton showers, iron initiated showers have more muons, which is readily seen by looking at the maximum of the profile\footnote{in this paper the index $\mu$ and $e.m.$ will be used to address to variables related with the MPD profile and the electromagnetic profile, respectively.}, $N^{\mu}_{max}$. Moreover, the height of the maximum fluctuates more for proton induced showers than for iron showers. This is a distinct feature from the electromagnetic profile that, having its $N^{e.m.}_{max}$ related to the shower energy, has essentially the same maximum value for both proton and iron shower, whereas the fluctuation level is of the order of 5\%.

\begin{figure}
\centering
\sidecaption
\includegraphics[width=0.45\textwidth,clip]{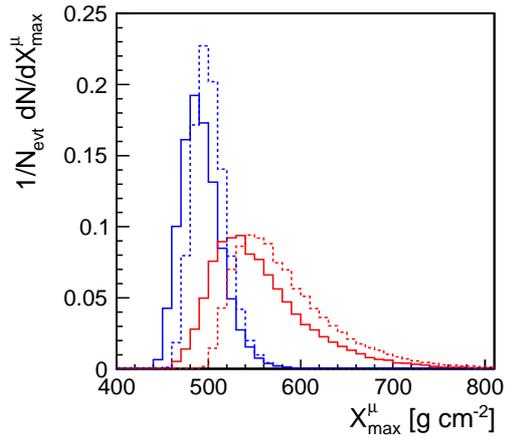}
\caption{$X^{\mu}_{max}$ distributions of muon production longitudinal profiles for different primaries and hadronic interaction models - the red lines are for proton and the blue ones for iron primaries. The distributions generated with \qII\ are shown as full lines while the ones generated with EPOS are displayed with dashed lines.}
\label{fig:xmax}       
\end{figure}

The second feature worth noticing is the depth of the shower maximum, \Xmumax, and the corresponding distribution which can be seen in more detail in Fig.~\ref{fig:xmax}. Here one can see that the behaviour of this observable follows the one found in the electromagnetic profile. The \Xmumax\ of proton induced showers are deeper and fluctuate more than in iron showers. This observable is sensitive to both primary mass composition and hadronic interaction physics, being the difference between proton and iron bigger than the difference for different hadronic models (\qII.03~\cite{QGSII1,QGSII2} results are displayed as full lines and EPOS1.99~\cite{EPOS} as dashed lines).

\begin{figure}
\centering
\sidecaption
\includegraphics[width=0.45\textwidth,clip]{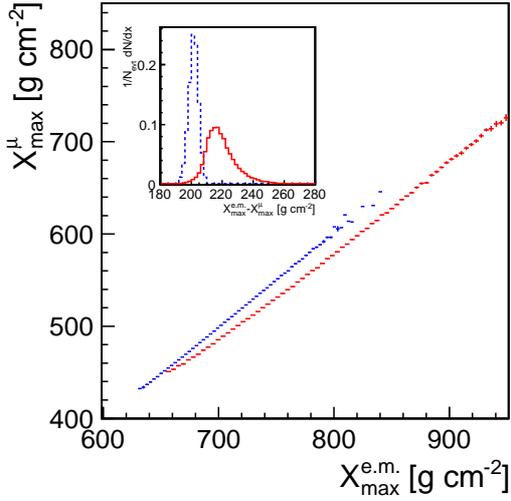}
\caption{Relation between muon production $X^{\mu}_{max}$ and electromagnetic $X^{e.m.}_{max}$, for $E = 10^{19}$ eV showers. The difference between the two $X_{max}$ are shown in the inset plots. The dependence on primary mass is shown (proton is the red (full) line and iron the blue (dashed) line, both generated with \qII).}
\label{fig:Xmaxem-mu}       
\end{figure}

An interesting correlation to look at is the one between the \Xmumax\ and \Xemmax, in an event-by-event basis. This profile is shown in Fig.~\ref{fig:Xmaxem-mu} for proton (in red) and iron (in blue) induced showers, both generated with \qII. The electromagnetic maximum is reached around $200~\rm{g~cm^{2}}$ later than the muonic profile maximum. On average, there is a correlation between the electromagnetic and the muonic \Xmax. It depends slightly on the primary mass composition and it was checked to be nearly independent of hadronic interaction models.

The difference between electromagnetic and muonic \Xmax, shown as an inset plot of Fig.~\ref{fig:Xmaxem-mu}, also changes between primaries $-$ by around $30~\rm{g~cm^{2}}$ in the average separation $-$ which means that although related, the two profiles give some independent information. This quantity, \Xemmax$-$\Xmumax, is very interesting as it is by construction independent of the first interaction point, which means that it should not be affected by sudden changes in the primary cross-section as suggested in~\cite{XsecBlackDisc}.

\section{Universality of the longitudinal profile shape}
\label{sec:usp}

\begin{figure}
\centering
\sidecaption
\includegraphics[width=0.45\textwidth,clip]{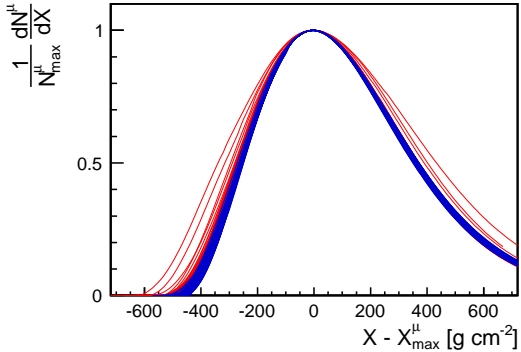}
\caption{Muon production shower profiles from proton (red) and iron (blue) primaries, in $(X',N')$ coordinates. The same showers used to build Fig. \ref{fig:prof} are used here.}
\label{fig:uspprof}       
\end{figure}

In Fig.~\ref{fig:uspprof} the profiles shown in Fig.~\ref{fig:prof} are expressed in $X^\prime \equiv X - X_{max}$ and $N^\prime \equiv N/N_{max}$. In these variables one can see that the obtained shape is rather universal, similarly to what happens to the electromagnetic profile.

Taking advantage of this universality it can be useful to use the average shape in order to determine the two main parameters from a fit with a reduced set of data. Moreover, there can be extra information on the average shape similar to what happens in the electromagnetic profile~\cite{USPV}. In fact, it was found that iron showers have a broader half mean width and the rise of the profile happens faster than for proton induced showers. The impact of different hadronic interaction models in the shape was found to be very small~\cite{USPVmuons}.

\begin{figure}
\centering
\sidecaption
\includegraphics[width=0.45\textwidth,clip]{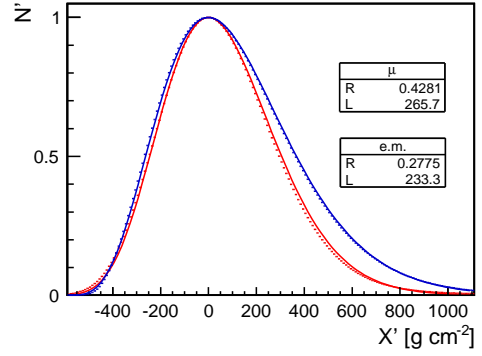}
\caption{Average shower profiles for proton primaries at E = $10^{19}$ eV, with \qII, in $(X',N')$ coordinates. Comparison between electromagnetic (in red) and muonic (in blue) shape features. The lines correspond to fits using a Gaisser-Hillas function (2 parameters). The fit results are given in the plot for the electromagnetic (e.m.) and muonic ($\mu$) profiles.}
\label{fig:profemdmuusp}       
\end{figure}

This quasi-universal shape is compared to the electromagnetic shape in Fig.~\ref{fig:profemdmuusp}. The MPD profile has a steeper growth and is more asymmetric, with respect to the shower maximum. Both profiles were fitted with a Gaisser$-$Hillas function, written in terms of a Gaussian width $L$ and a parameter $R$, which is related with the asymmetry of the shower with respect to the shower maximum~\cite{USPV}, such that,

\begin{equation}
        N' = \left( 1 + \frac{ R X' }{ L} \right)^{R^{-2}} \exp \left( - \frac{ X' }{ LR } \right).
        \label{USPVform}
\end{equation}

We conclude that the MPD profile can be  well described with the same function as used for the electromagnetic  profile. In fact, it is worth noting that the full profile description is better achieved for the MPD profile. This is due to the fact that the energy deposit profile (the profile that is seen by the fluorescence detector and is intimately related with the electromagnetic shower component) contains in the end-tail a non-negligible contribution from muons that is not accounted by the Gaisser$-$Hillas. Nevertheless, it is interesting to note that both distributions, having different development mechanisms, share the same kind of structure.

\begin{figure}
\centering
\sidecaption
\includegraphics[width=0.45\textwidth,clip]{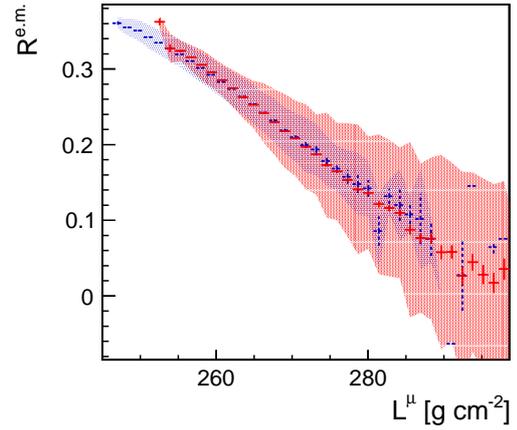}
\caption{Correlation between the electromagnetic profile shape (characterized by $R^{e.m.}$) and the muon production profile shape (represented by $L^{\mu}$). The results are shown for proton (red/full) and iron (blue/dashed) induced showers,  at E = $10^{19}$ eV, and with \qII. The shaded area shows the region within one sigma.}
\label{fig:shapecorr}       
\end{figure}

Through the study of the shape parameters ($R,L$) we were able to conclude that, similarly to what happened with the electromagnetic profile, one of the parameters retains almost all information on the type of primary allowing the other one to be fixed. However, while in the e.m. case it was $L^{e.m.}$ that, being related to the shower energy, could be fixed, in the MPD profile it is $L^{\mu}$ that shows more capabilities to distinguish between proton and iron induced showers. Note that $L^{\mu}$ is proportional to the integral of the MPD profile which attains information about the total number of muons produced during the shower development.

Having a single shape variable to characterise each profile, $L^{\mu}$ and $R^{e.m.}$, the correlation between the shape of the MPD and the electromagnetic profile can be easily drawn. The relation between the shapes of the two profiles for each individual event can be seen in fig. \ref{fig:shapecorr}. Here, $(L^{e.m.}, R^{\mu})$ were fixed to its corresponding average values of the combined proton and iron distributions. The correlation is rather strong in the most populated region of $(R^{e.m.}, L^{\mu})$. Moreover, it is almost independent of the primary mass composition, making it very useful for hybrid analysis. Indeed, whenever one of the profiles is measured accurately a prediction of the other profile shape can be established allowing, in this way to perform consistency tests to the shower description.

\section{ Constraining Hadronic Interaction Models}
\label{sec:umbrella}

Finally, in this section we want to point out some strategies on how to disentangle the mass composition from the hadronic interaction models uncertainty. For this, it is useful to introduce the two-dimensional plots where the two first momenta of \Xemmax\ distribution, namely \avXemmax\ and \RMSXemmax, are plotted for each mass composition scenario at a fixed energy~\cite{Linsley,UngerKampert}. In these kind of plots each point represents a possible scenario and by considering the possible primary mass combinations we are, in a sense, probing the \emph{phase space} for a given hadronic interaction model.

\begin{figure}
\centering
\sidecaption
\includegraphics[width=0.45\textwidth,clip]{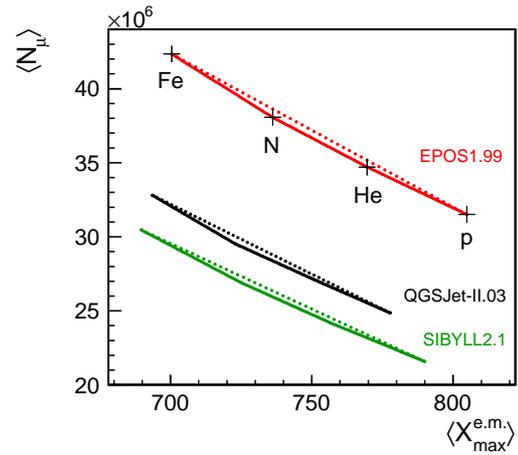}
\caption{Average number of muons \emph{versus} average electromagnetic \Xemmax\ for E = $10^{19}$ eV showers. Each single point in the plot represents a primary mass composition scenario. The results are shown for three distinct hadronic interaction models:  \qII.03 (in black); EPOS1.99 (in red) and SIBYLL2.1 (in green). For EPOS the pure mass composition scenarios are marked with black crosses and the label below identifies the specie.}
\label{fig:umbrella_nmu_xmax}       
\end{figure}

Here, we want to extend this concept and combine two observables, that are sensitive to both the hadronic interaction models and primary mass composition: the average total number of muons produced in the shower\footnote{this quantity is obtained performing the integral of the MPD profile up to the maximum. In this way, we obtain a quantity which is proportional to the total number of muons in the shower without any ground effects. For a fixed zenith angle the number of muons at ground gives similar results.}, $\left< N_{\mu} \right>$, and the average electromagnetic shower maximum depth, \avXemmax. The resulting plot is shown in Fig.~\ref{fig:umbrella_nmu_xmax} for three distinct hadronic interaction models: \qII.03 (in black); EPOS1.99 (in red) and SIBYLL2.1~\cite{SIBYLL21} (in green). Let us now evaluate the results of a single model, namely EPOS. The most extreme point at the left represents a scenario with pure iron as mass composition, while the point at the right is for a scenario where there are only protons. The pure helium and nitrogen scenario are also marked for the case of EPOS. The remaining red dots are the results for bimodal mass composition scenario from one pure element to another. Inside of this contour are all the remaining complex mass composition combination amongst the four species: p, He, N, Fe.

The most interesting feature of this plot is that while the problem is degenerate when $\left< N_{\mu} \right>$ and \avXemmax\ are interpreted alone, here the hadronic interaction models can be distinguished independently of any mass composition scenario, provided, of course, that the experimental resolutions are better than the separation between models. Moreover, these kind of analyses are important to constrain the allowed \emph{phase space} for hadronic interaction models, increasing in this way our knowledge about the shower. 

\section{Conclusions}
\label{sec:conclusions}
The muon production longitudinal profile of air showers gives new primary mass composition variables which are fairly independent of the high energy hadronic interaction  model: the $X^{\mu}_{max}$ and the shape variable $L^{\mu}$. The normalisation of the profile gives also access to the total number of produced muons, which is known to be an important variable, both for primary composition and high energy hadronic interaction model studies.

The MPD profile can be described using the parameterisation that is used for the electromagnetic profile. Moreover, its shape displays an universal behaviour when expressed in terms of $X^\prime$ and $N^\prime$.

Combining the information of the muonic longitudinal profile at production with the electromagnetic profile will give rise to extra variables, sensitive to the shower development characteristics. These will lead to a more precise understanding of the shower development. In fact, we have shown that already some constrains on hadronic interaction models can be applied, independently of any mass composition scenarios, through the combination of the average electromagnetic shower maximum depth and the average number of muons produced in the shower.

With the higher number of available observables, using profiles which are independent of the detection conditions and more directly related to the hadronic cascade, cosmic rays become increasingly useful for the study of particle physics at the highest energies.

\section*{Acknowledgments}
We would like to thank J. Alvarez-Mu\~niz for careful reading the manuscript. This work is partially funded by Funda\c{c}\~{a}o para a Ci\^{e}ncia e Tecnologia (CERN/FP11633/2010 and SFRH/BPD/73270/2010), and fundings of MCTES through POPH-QREN-Tipologia 4.2, Portugal, and European Social Fund.

\bibliography{Bib-hX}
%

\end{document}